\input{aipcheck}

\documentclass[
  ,final            
] {aipproc}

\layoutstyle{6x9}

\begin{document}

\title{Spin Identification of the Randall-Sundrum Graviton at the LHC}

\classification{12.60.-i, 11.10.Kk, 12.60.Cn} \keywords      {LHC,
graviton, spin}

\author{P.\ Osland}{
address={Department of Physics and Technology, University of
Bergen, N-5020 Bergen, Norway} }

\author{A.\ A.\ Pankov}{
address={The Abdus Salam ICTP Affiliated Centre, Technical
University of Gomel, 246746 Gomel, Belarus} }

\author{A.\ V.\ Tsytrinov}{
address={The Abdus Salam ICTP Affiliated Centre, Technical
University of Gomel, 246746 Gomel, Belarus} }

\author{N.\ Paver}{
address={University of Trieste and INFN-Trieste Section, 34100
Trieste, Italy} }

\begin{abstract}
Using as basic observable an angular-integrated asymmetry to be
measured in Drell-Yan lepton-pair production at the LHC, we
discuss the identification reach on the spin-2 of the lowest-lying
Randall-Sundrum resonance predicted by gravity in one warped extra
dimension, against the spin-1 and spin-0 hypotheses. Numerical
results indicate that, depending on the graviton coupling strength
to the standard model particles, such a spin-2 identification can
extend up to mass scales of 1.0--1.6~TeV and 2.4--3.2~TeV for LHC
integrated luminosities of 10 and 100~${\rm fb}^{-1}$,
respectively.
\end{abstract}

\maketitle


\section{Introduction}
Heavy quantum states, with masses $M\gg M_{W,Z}$, are generally
predicted by new physics (NP) models. If the masses are in the TeV
range, such non-standard objects could be directly revealed as
peaks, or resonances, in the cross sections for reactions among
standard model (SM) particles at supercolliders.
\par
The {\it discovery reach} represents the upper limit of the mass
range where a peak can be observed experimentally. It depends,
among other things, on the collider energy and the expected
statistics, and determines an accessible region for the NP model
parameters. However, once a peak is observed, the determination of
the underlying model against others, potentially giving the same
mass and number of events, is needed. Accordingly, for any model,
the {\it identification reach} defines the upper limit of the mass
range where the source of the peak can be determined or,
equivalently, the competitor models can be excluded for all values
of their parameters. Clearly, the identification reach is expected
to select a subdomain of the model parameters accessible to
discovery.
\par
The resonance spin represents a powerful discriminating observable
in this regard. Popular examples of NP scenarios that can produce
(narrow) peaks in cross sections with the same mass and number of
events, and can be discriminated by a spin analysis are: {\it i)}
Models of gravity in extra spatial dimensions (spin-2); {\it ii)}
Models with heavy neutral gauge bosons $Z^\prime$ (spin-1); and
{\it iii)} SUSY models with R-parity breaking sneutrino couplings
(spin-0). Here, we discuss the identification reach on the
{lowest-lying}, {spin-2}, Randall-Sundrum (RS) graviton
resonance~\cite{Randall:1999ee}, against {\it both} the {spin-1}
and {spin-0} hypotheses, that can be obtained from the
experimental study of the inclusive dilepton production process at
LHC ($l=e,\mu$):
\begin{equation}
p+p\to l^+l^-+X. \label{proc}
\end{equation}
While the total resonant cross section, integrated over the
dilepton invariant mass under the peak at $M=M_G$, determines the
number of events, hence the discovery reach, for the assessment of
the spin-2 identification reach we adopt as basic observable a
specific angular-integrated asymmetry, $A_{\rm CE}$, at
$M=M_G$~\cite{Osland:2008sy}. The angle is that between the final
lepton and the initial quark or gluon in the dilepton
center-of-mass frame. This asymmetry has the built-in feature of
directly disentangling the spin-1 from other spin
hypotheses~\cite{Dvergsnes:2004tw,Osland:2003fn} and, being a
ratio of cross sections, should be less affected by systematic
uncertainties than other observables. Earlier attempts to
discriminate spin-2 from spin-1, based on `absolute' angular
differential distributions, have been presented, {\it e.g.}, in
Refs.~\cite{Allanach:2000nr,Cousins:2005pq} and in the
experimental Ref.~\cite{Abulencia:2005nf}.
\section{New physics models}
\leftline{\bf RS model of gravity in extra dimensions} This
scenario is a candidate solution to the gauge hierarchy problem,
and its simplest version is based on one compactified `warped'
spatial extra dimension and a two-brane setup. The SM particles
are localized to the TeV brane, gravity can propagate in the full
5-dimensional bulk and in particular, on the Planck brane, has an
effective scale determined by ${\overline M}_{\rm
Pl}=1/{\sqrt{8\pi G_{\rm N}}}=2.44\times 10^{18}\, {\rm GeV}$. On
the TeV brane the gravity scale $\Lambda_\pi$ is suppressed by the
exponential `warp' factor, $\Lambda_\pi={\overline M}_{\rm
Pl}\,{\exp{(-\pi k R_c)}}$, with $k\sim {\overline M}_{\rm Pl}$
the 5-dimensional curvature and $R_c$ the compactification radius.
Boundary conditions at the branes determine a tower of spin-2
graviton  resonances $G^{(n)}$ ($n\ge 1$). Their predicted mass
spectrum is $M_n=M_1x_n/x_1$ with $M_1$ the lowest resonance mass
and $x_n$ the roots of the Bessel function $J_1(x_n)=0$. Their TeV
brane couplings to the SM particles are given by
\begin{equation}
-{\cal L}_{\rm TeV}=\left[\frac{1}{{\overline M}_{\rm
Pl}}G_{\mu\nu}^{(0)}(x)+
\frac{1}{\Lambda_\pi}\sum_{n=1}^{\infty}G_{\mu\nu}^{(n)}(x)\right]T^{\mu\nu}(x),
\label{interaction}
\end{equation}
where $T^{\mu\nu}$ is the energy-momentum tensor and $G^{(0)}$
denotes the zero-mode, ordinary, graviton. For $kR_c\simeq 12$,
$\Lambda_\pi$ as well as the masses  $M_n$ (through the relation
$M_1=\Lambda_\pi kx_1/{\overline M}_{\rm Pl}$) are ${\cal O}(\rm
TeV)$. This opens up the interesting possibility of observing
gravity effects at colliders, in particular to reveal the graviton
excitation exchange in the process (\ref{proc}). The contributing,
tree level, partonic processes
\begin{equation}
q{\bar q}\to G\to l^+l^- \qquad\quad {\rm and} \qquad\quad gg\to
G\to l^+l^- \label{graviton}
\end{equation}
should yield cross-section peaks at the invariant dilepton mass
$M=M_n$, with characteristic angular
distributions~\cite{Han:1998sg}.
\par
The RS model thus depends on two independent parameters, that can
be chosen as $M_G\equiv M_1$ and the universal `coupling'
$c=k/{\overline M}_{\rm Pl}$ (in which case $\Lambda_\pi$ is a
derived parameter). Theoretically `natural' limits are $0.01\leq c
\leq 0.1$ and $\Lambda_\pi<10~{\rm TeV}$~\cite{Davoudiasl:2000jd}.
Current 95\% C.L. experimental lower limits on $M_G$, from the
Tevatron collider, range from 300 GeV ($c=0.01$) to 900 GeV
($c=0.1$)~\cite{Abazov:2007ra}. One should notice that the
unevenly spaced spectrum could be distinctive of the model by
itself. However, in practice, due to the large masses involved,
only the first RS resonance might be accessible at LHC, so that
the spin-2 determination should be a necessary test of the RS
model.
\medskip
\par\noindent
{\bf Heavy neutral gauge bosons}
\par\noindent
Turning to spin-1 exchanges in the process (\ref{proc}),
$Z^\prime$s generally occur in electroweak models based on
extended gauge symmetries. The leading-order partonic process,
$q{\bar q} \to Z^\prime \to l^+l^-$, should show up as a peak in
the dilepton invariant mass distribution at $M=M_{Z^\prime}$ with,
in this case, the {\it same} angular distribution as the SM
$\gamma$ and $Z$ exchanges. Besides the mass $M_{Z^\prime}$, the
model parameters are the vector and axial-vector $Z^\prime$
couplings to quarks and leptons. Popular scenarios are the cases
where such couplings are specified theoretically: the list
includes $Z^\prime_\chi$, $Z^\prime_\psi$, $Z^\prime_\eta$,
$Z^\prime_{\rm LR}$, $Z^\prime_{\rm ALR}$ models, and the
`sequential' $Z^\prime_{\rm SSM}$ model with the same couplings as
the SM. Details can be found, {\it e.g.}, in the recent
Ref.~\cite{Langacker:2008yv}. It turns out that, at the assumed
LHC luminosity, the spin-2 RS resonance can be distinguished from
the ALR and SSM spin-1 scenarios already at the level of signal
events~\cite{Osland:2008sy}. For the other $Z^\prime$ models,
there are `confusion regions' in the parameter spaces where spin-2
and spin-1 exchanges give rise to the same peaks and number of
events, and therefore can be distinguished by a spin analysis
only. Current experimental, model-dependent, lower limits on
$M_{Z^\prime}$, from the Tevatron collider, are in the range
500-900 GeV~\cite{Tev:2007sb}.
\medskip
\par\noindent
{\bf Sneutrino exchange}
\par\noindent
In SUSY theories with R-parity breaking, sparticles can be
exchanged in the process (\ref{proc}) and appear as peaks in the
dilepton invariant mass. This is the case of the spin-0 sneutrino
formation by quark-antiquark annihilation, followed by leptonic
decay \cite{Kalinowski:1997bc}: $q {\bar q}\to {\tilde\nu} \to
l^+l^-$. At the peak in the dilepton invariant mass,
$M=M_{\tilde\nu}$, the spin-0 character implies a flat angular
distribution. Basically, the cross section in this model depends
on $M_{\tilde\nu}$ and on the product $X=(\lambda^\prime)^2B_l$,
where $\lambda^\prime$ is the R-parity breaking sneutrino coupling
to $d\bar d$ and $B_l$ the sneutrino leptonic branching ratio.
Current constraints on $X$ are very loose, and there exists an
extended domain where $\tilde\nu$ production can mimic RS
resonance formation (same mass and number of events under the
peak), for details see~\cite{Osland:2008sy}.
\section{${\bf{A}}_{\bf CE}$ asymmetry and spin-2 RS graviton identification}
With $z=\cos\theta_{\rm cm}$ and $R=G,V,S$ denoting the spin-2,
spin-1 and spin-0 hypotheses, respectively, we define the evenly
integrated center-edge asymmetry:
\begin{equation}
\label{ace}
A_{\rm{CE}}(M_R)=\frac{\sigma_{\rm{CE}}(R_{ll})}{\sigma(R_{ll})}\quad{\rm
with} \quad \sigma_{\rm{CE}}(R_{ll}) \equiv
\left[\int_{-z^*}^{z^*} - \left(\int_{-z_{\rm cut}}^{-z^*}
+\int_{z^*}^{z_{\rm cut}}\right)\right] \frac{{\rm d}
\sigma(R_{ll})}{{\rm d} z}\, {\rm d} z.
\end{equation}
In (\ref{ace}): $0<z^*<{z_{\rm cut}}$ is a priori free, and
defines the separation between the ``center'' and the ``edge''
angular regions; $\vert z\vert < z_{\rm cut}$ accounts for
detector angular acceptance; cross sections are integrated over
the lepton-pair rapidity and over a bin in the lepton-pair
invariant mass $M$ centered at  the peak $M=M_R$ and with size
$\Delta M$ appropriate to account for the detector resolution,
see, {\it e.g.}, Ref.~\cite{Atlas}. To a very good approximation,
the explicit $z^*$ dependencies of $A_{\rm CE}$ are, for the three
spin-hypotheses:
\begin{equation}\label{ACE2}
A_{\rm CE}^{G} =\epsilon_q^{\rm SM}\,A_{\rm CE}^{V} +
\epsilon_q^G\left[2\,{z^*}^5+\frac{5}{2}\,z^*(1-{z^*}^2)-1\right]
+ \epsilon_g^G\left[\frac{1}{2}\,{z^*}(5-{z^*}^4)-1\right],
\end{equation}
\begin{equation}\label{ACE1-ACE0}
A_{\rm CE}^{V} \equiv A_{\rm CE}^{\rm SM}
=\frac{1}{2}\,z^*({z^*}^2+3)-1, \quad
A_{\rm CE}^{S} = \epsilon_q^{\rm SM}\,A_{\rm CE}^{V}
+\epsilon_q^{S}\,(2\,z^*-1).
\end{equation}
In (\ref{ACE2}), $\epsilon_q^G$, $\epsilon_g^G$ and
$\epsilon_q^{\rm SM}$ are the fractions of resonant events for
$q\bar q,gg\to G\to l^+l^-$ and SM background, respectively, with
$\epsilon_q^G+\epsilon_g^G+ \epsilon_q^{\rm SM}=1$. They are
determined, as functions of $M$, by the overlaps of parton
distribution functions, for which  we choose the CTEQ6
ones~\cite{Pumplin:2002vw}. Analogous definitions hold for
Eq.~(\ref{ACE1-ACE0}). Strictly,
Eqs.~(\ref{ACE2})-(\ref{ACE1-ACE0}) are exact in the limit $z_{\rm
cut}=1$, whereas we will impose $z_{\rm cut}=0.987$: the
difference is numerically negligible at the `optimal' values
$z^*\simeq 0.5$ used in the subsequent analysis. The numerical
results presented here are obtained from `full' calculations with
foreseen experimental cuts, such as lepton pseudorapidity
$\vert\eta_l\vert<2.5$ and transverse momenta $p_{T,l}>20\, {\rm
GeV}$. Also, a (perhaps optimistic) lepton identification
efficiency of 90\% has been assumed to evaluate the statistics.
\par
One should notice, in Eq.~(\ref{ACE1-ACE0}), that $A_{\rm
CE}^{V}\equiv A_{\rm CE}^{\rm SM}$ and, therefore, deviations of
$A_{\rm CE}$ from the SM predictions definitely signal NP
exchanges different from spin-1 models.
\par
We now suppose that a peak is discovered in the dilepton mass
distribution for the process (\ref{proc}) at $M=M_R$, and make the
hypothesis that it is consistent with a spin-2 RS graviton
resonance (in which case, $M_R$ must be identified as $M_G$). To
assess the level at which the spin-1 {\it and} spin-0 hypotheses
can be excluded as competing sources of the peak with the same
number of events, hence the spin-2 hypothesis being established,
one can consider the deviations:
\begin{equation}
\Delta A_{\rm CE}^V=A_{\rm CE}^G-A_{\rm CE}^V \qquad{\rm
and}\qquad \Delta A_{\rm CE}^S=A_{\rm CE}^G-A_{\rm CE}^S.
\label{deltaGSV}
\end{equation}
\par
As an example, Fig.~7 of Ref.~\cite{Osland:2008sy} shows $A_{\rm
CE}$ {\it vs} $z^*$ for resonances with different spins, the same
mass $M_R=1.6~{\rm TeV}$ and the same number of events, $c=0.01$
for the RS exchange, and LHC luminosity ${\cal L}_{\rm int}=100\,
{\rm fb}^{-1}$ (left panel). The right panel of that figure shows
the corresponding deviations (\ref{deltaGSV}). The vertical bars
attached to the dot-dashed line representing the RS model, give
2$\sigma$ statistical uncertainties on the model itself. The
figure suggests that, indeed, at the assumed LHC luminosity, the
spin-2 RS graviton with $M_G=1.6~{\rm TeV}$ and $c=0.01$ can be
discriminated from the other spin hypotheses by means of $A_{\rm
CE}$ at $z^*\simeq 0.5$.
\par One can systematically generalize this example, and look at the domain in the RS parameter plane ($M_G,c$) where a peak can be {\it identified} as originating from spin-2 RS exchange, with the spin-1 and spin-0 hypotheses excluded. This domain will represent the searched for {\it identification reach} on the RS graviton resonance. For this purpose, one can adopt a simple-minded $\chi^2$ criterion, where the $\chi^2$ functions are defined as $\chi^2=\left[\Delta A_{\rm CE}/\delta A_{\rm CE}\right]^2$, with $\Delta A_{\rm CE}$ given in (\ref{deltaGSV}), and $\delta A_{\rm CE}$ the corresponding expected statistical uncertainty pertaining to the RS model. The conditions $\chi^2>\chi_{\rm C.L.}^2$ determine the ranges in ($M_G,c$) where the spin-0 and spin-1 hypotheses can be excluded to a given confidence level. The maximum sensitivity of $A_{\rm CE}$ to the spin-2 RS resonance parameters is generally achieved for $z^*=0.5$.
\begin{figure}
 \includegraphics[height=.255\textheight]{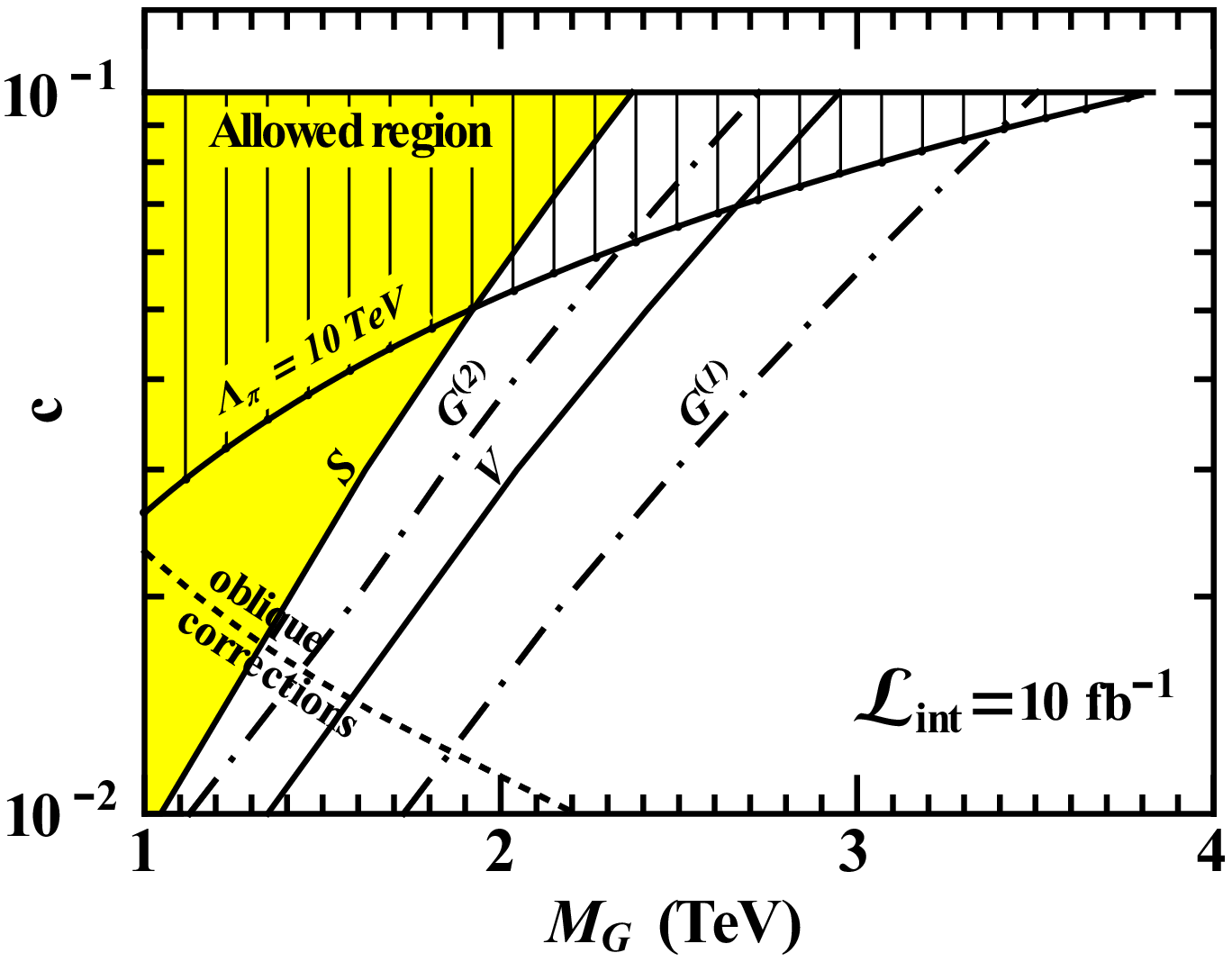} \hspace{2mm}
  \includegraphics[height=.255\textheight]{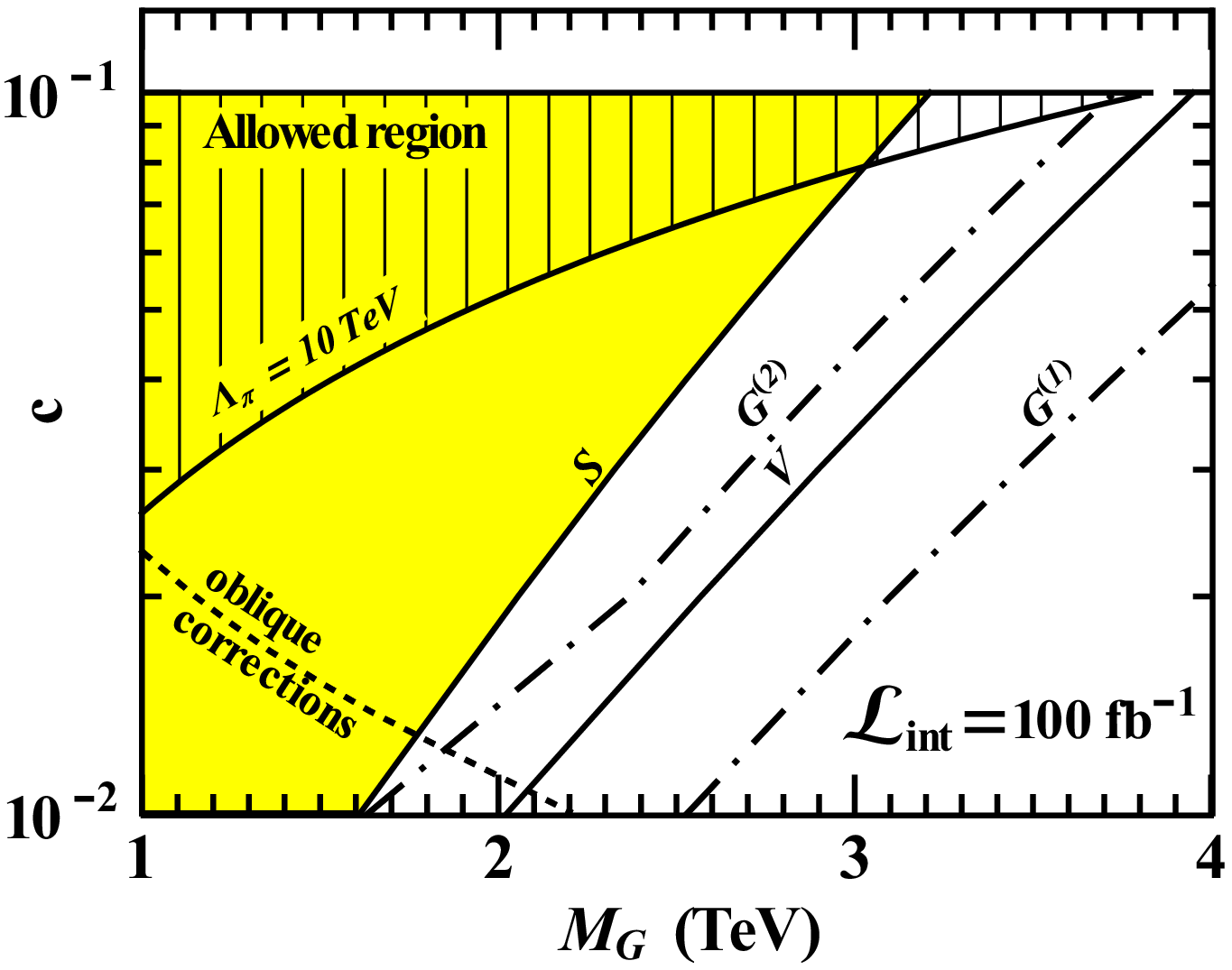}
 \caption{Discovery and Identification ranges as defined in the text.}
\end{figure}
\par
Figure~1 shows the resulting allowed domain in the RS parameter
plane for the spin-2 discrimination at 95\% C.L. {\it vs.} the
allowed domain for discovery at 5$\sigma$,  for two values of LHC
integrated luminosity, and the channels $l=e,\mu$
combined~\cite{Osland:2008sy}. Theoretically suggested bounds on
$c$ and $\Lambda_\pi$ are taken into account. The meaning of the
lines in the two panels are as follows: the lowest RS resonance
can be discovered if its representative point ($M_G,c$) lies to
the left of the line ``$G^{(1)}$''; in the region to the left of
curve ``$V$'', of course included in the preceding one, the RS
resonance spin-1 can be excluded, whereas the spin-0 hypothesis is
still open; finally, if the observed resonance has the
representative point in the domain to the left of the line
``$S$'', also the spin-0 hypothesis can be excluded. Accordingly,
this is assumed to represent the searched for domain allowing
spin-2 identification. Of course, these statements should be
understood in a statistical sense, as specified by the confidence
level.
\par
As regards the mass ranges for discovery and identification of the
RS graviton resonance, the results in Fig.~1 can be summarized
as in Tab.~1. One should notice that, in Fig.~1, the line
``$S$'' always lies to the left of the line ``$V$''. This reflects
the (general) fact that, as seen from
Eqs.~(\ref{ACE2})--(\ref{deltaGSV}), $\Delta A_{\rm CE}^V > \Delta
A_{\rm CE}^S$ for all $z^*$. Thus, if one were able to exclude
spin-0, the exclusion of the whole class of spin-1 models would be
automatically implied in a model-independent way. Conversely, the
request of excluding the spin-0 hypothesis substantially reduces
the extension of the parameter domain allowed by the weaker
condition of only discriminating spin-2 from spin-1.
\par
The somewhat `low' values of $M_G$ allowed to spin-0 exclusion
suggest to look at the possibility of discovering the next RS
resonance, $G^{(2)}$, in addition to $G\equiv G^{(1)}$, recalling
that the ratio of masses is in the model a predicted number,
$M_G/M^{(2)}=x_1/x_2$. Indeed, such 5$\sigma$ discovery of
$G^{(2)}$ turns out to be possible, with ${\cal L}_{\rm int}=10\,
{\rm fb}^{-1}$ for $M_G < 1.1\, {\rm TeV}$ (2.7 TeV) at  $c=0.01$
($c=0.1$) and, with ${\cal L}_{\rm int}=100\, {\rm fb}^{-1}$, for
$M_G<1.6~{\rm TeV}$ (3.7 TeV) at $c=0.01$ ($c=0.1$).
Correspondingly, for a lowest resonance $G$ in the ($M_G,c$)
domain to the left of the line ``$G^{(2)}$'' in Fig.~1, also the
higher graviton excitation with $n=2$ can be discovered. One can
see, therefore, that to the left of the line ``$S$'' the spin-2 of
the lowest-lying RS resonance can be established and, in addition,
also the characteristic RS mass spectrum can be tested by the
discovery of the higher resonance, so that the model would be
doubly clinched.
\par
One can notice from Fig.~1 the dramatic role in RS graviton
searches of the bound $\Lambda_\pi\leq 10\, {\rm TeV}$,
theoretically motivated by the need of not creating additional
hierarchies in the model: taken literally, it would imply that, at
the high luminosity of 100 ${\rm fb}^{-1}$, the mass spectrum test
should be feasible in the full discovery domain. However, in
practice, such bound should be considered in a qualitative sense,
as is the case for the indicative limits from the fit to oblique
parameters, taken from~\cite{Davoudiasl:2000jd,han2000}.
\begin{table}
\begin{tabular}{lrrrr}
\hline
 & \tablehead{2}{r}{b}{Discovery}
 & \tablehead{2}{r}{b}{Identification} \\
\hline
${\cal L}_{\rm int}$ & $c=0.01$ &  $c=0.1$ &  $c=0.01$ &  $c=0.1$ \\
\hline
$10~{\rm fb}^{-1}$ & 1.7~TeV & 3.5~TeV & 1.1~TeV    & 2.4~TeV\\
$100~{\rm fb}^{-1}$ & 2.5~TeV & 4.6~TeV & 1.6~TeV  & 3.2~TeV\\
\hline
\end{tabular}
\caption{Discovery and Identification reach [TeV]} \label{tab:a}
\end{table}
\par
We may conclude by mentioning, as another RS resonance selective
process, the inclusive diphoton production $p+p \to G\to
\gamma\gamma + X$. Indeed, spin-1 could be excluded directly by
$V\not\to\gamma\gamma$, leaving only the spin-2 and spin-0
hypotheses, and the RS model could be strongly tested by
measurement of the ratio $Br(G\to\gamma\gamma)/Br(G\to
l^+l^-)$~\cite{Randall:2008xg}. Currently, only the diphoton
invariant mass distribution has been studied experimentally, but
angular analysis should be possible, as mentioned in
Refs.~\cite{Allanach:2000nr}, notwithstanding the dominance of the
partonic process $gg\to G\to \gamma\gamma$, strongly peaked near
$z=\pm 1$ and potentially affected by the initial bremsstrahlung
background. It could be interesting to attempt the application of
the asymmetry $A_{\rm CE}$ to this process also.

\end{document}